\documentclass[twocolumn,aps,prl,showpacs,floatfix,superscriptaddress]{revtex4}
\usepackage{amsmath}
\usepackage{amsfonts}
\usepackage{amssymb}
\usepackage{bm}
\usepackage{color}
\usepackage{graphicx}

\begin{document}

\title{ Universal intermittent  properties of particle trajectories in
  highly turbulent flows } 

\author{International Collaboration for Turbulence Research}\noaffiliation{}

\author{A. Arn\`eodo}
 \affiliation{Laboratoire de Physique de l'\'Ecole Normale Sup\'erieure
de Lyon,  46 all\'ee d'Italie F-69007 Lyon,
France} 

\author{R. Benzi} \affiliation{University of Tor Vergata and INFN, via
della Ricerca Scientifica 1, 00133 Rome, Italy}

\author{J. Berg} \affiliation{Wind Energy Department Ris\o\, National
Laboratory - DTU Denmark}

\author{L. Biferale$^*$} \affiliation{University of Tor Vergata and
  INFN, Via della Ricerca Scientifica 1, 00133 Rome, Italy}

\author{E. Bodenschatz}\affiliation{Max Planck Institute for Dynamics
 and Self-Organization, Am Fassberg 17, D-37077 Goettingen, Germany}

\author{A. Busse} \affiliation{Max Planck Institute for Plasma
Physics, Garching, Germany}

\author{E. Calzavarini}\affiliation{Department of Applied Physics,
University of Twente, 7500 AE Enschede, The Netherlands}

\author{B. Castaing} 
 \affiliation{Laboratoire de Physique de l'\'Ecole Normale Sup\'erieure
de Lyon,  46 all\'ee d'Italie F-69007 Lyon,
France} 

\author{M. Cencini$^*$} \affiliation{INFM-CNR, SMC Dipartimento di Fisica,
Universit\`a di Roma ``La Sapienza'', Piazzale A. Moro, 2 I-00185
Rome, Italy}

\author{L. Chevillard}
 \affiliation{Laboratoire de Physique de l'\'Ecole Normale Sup\'erieure
de Lyon,  46 all\'ee d'Italie F-69007 Lyon,
France} 

\author{R. T. Fisher} \affiliation{Department of Astronomy and
Astrophysics, The University of Chicago, Chicago, IL 60637, USA}

\author{R. Grauer} \affiliation{Institute for Theoretical Physics I,
Ruhr-University Bochum, Germany}

\author{H. Homann} \affiliation{Institute for Theoretical Physics I,
Ruhr-University Bochum, Germany}

\author{D. Lamb} \affiliation{Department of Astronomy and
Astrophysics, The University of Chicago, Chicago, IL 60637, USA}

\author{A. S. Lanotte$^*$} \affiliation{CNR-ISAC, Via Fosso del
Cavaliere 100, 00133 Rome and INFN, Sezione di Lecce, 73100 Lecce, Italy}

\author{E. L\'ev\`eque} \affiliation{Laboratoire de Physique de
  l'\'Ecole Normale Sup\'erieure de Lyon, 46 all\'ee d'Italie F-69007
  Lyon, France}

\author{B. L\"uthi} \affiliation{IFU, ETH Z\"urich, Switzerland}

\author{J. Mann}
\affiliation{Wind Energy Department Ris\o\, National Laboratory - DTU
Denmark} 

\author{N. Mordant} \affiliation{Laboratoire de Physique Statistique
  de l'Ecole Normal Sup\'erieure CNRS 24 Rue Lhomond, 75231 Paris Cedex 05, France }

\author{W.-C. M\"uller}\affiliation{Max Planck Institute for Plasma
Physics, Garching, Germany} 

\author{S. Ott} \affiliation{Wind Energy
Department Ris\o\, National Laboratory - DTU Denmark}

\author{N. T. Ouellette} \affiliation{Haverford College, 370 Lancaster Avenue Haverford,
PA 19041, USA}

\author{J.-F. Pinton}
 \affiliation{Laboratoire de Physique de l'\'Ecole Normale Sup\'erieure
de Lyon,  46 all\'ee d'Italie F-69007 Lyon,
France} 

\author{S. B.  Pope}
\affiliation{Sibley School of Mechanical and Aerospace Engineering 254 Upson Hall, Cornell University, Ithaca, NY 14853-7501 USA} 

\author{S. G. Roux} 
 \affiliation{Laboratoire de Physique de l'\'Ecole Normale Sup\'erieure
de Lyon,  46 all\'ee d'Italie F-69007 Lyon,
France} 

\author{F. Toschi$^*$} \affiliation{Istituto per le Applicazioni del
Calcolo CNR, Viale del Policlinico 137, 00161 Roma, Italy}
\affiliation{INFN, Sezione di Ferrara,
  Via G. Saragat 1, 44100 Ferrara, Italy}

\author{H. Xu} \affiliation{Max Planck Institute for Dynamics and
Self-Organization, Am Fassberg 17, D-37077 Goettingen, Germany}

\author{P. K. Yeung} \affiliation{School of Aerospace Engineering, Georgia Institute of Technology,
279 Ferst Drive Atlanta, GA 30332-0150, USA}

\begin{abstract}

We present a collection of eight data sets, from state-of-the-art
experiments and numerical simulations on turbulent velocity statistics
along particle trajectories obtained in different flows with Reynolds
numbers in the range $R_\lambda \in [120:740]$.  Lagrangian structure
functions from all data sets are found to collapse onto each other on
a wide range of time lags, pointing towards the existence of a
universal behaviour, within present statistical convergence, and
calling for a unified theoretical description.  Parisi-Frisch
Multifractal theory, suitable extended to the dissipative scales and
to the Lagrangian domain, is found to capture intermittency of
velocity statistics over the whole three decades of temporal scales
here investigated.

\end{abstract}

\date{}

\maketitle 

 Understanding the statistical properties of particle tracers advected
 by turbulent flows is a challenging theoretical and experimental
 problem \cite{FS.06,LPVCAB.01}. It is a key ingredient for the
 development of stochastic models \cite{Po.00,lamorgese07}, in such
 diverse contexts as turbulent combustion, industrial mixing,
 pollutant dispersion and cloud formation \cite{Shaw.03}. The main
 difficulty of Lagrangian investigations, following particle
 trajectories, stems from the necessity to resolve the wide range of
 time scales driving different particle behaviours: from the longest,
 $T_L$, given by the stirring mechanism, to the shortest
 $\tau_{\eta}$, typical of viscous dissipation. Indeed the ratio,
 $T_L/\tau_{\eta} \sim R_{\lambda}$, grows with the Taylor Reynolds
 number, $R_{\lambda}$, that varies up to few thousands in laboratory
 flows. Some aspects of Lagrangian statistics have been experimentally
 measured: particle accelerations~\cite{LPVCAB.01}, velocity
 fluctuations in the inertial range~\cite{MMMP.01,X06} and
 two-particle dispersion~\cite{berg,B06}.  Others, connected to the
 entire range of motions, have long been restricted to numerical
 simulations \cite{BBCLT.05,MLP04,pkyeung,mueller,BBCDLT.04}. A
 fundamental open question is connected to intermittency, i.e.  the
 observed strong deviations from Gaussian statistics, becoming larger
 and larger when considering fluctuations at smaller and smaller
 scales. Besides, the dependency of velocity statistics at various
 temporal scales on large scale forcing and boundary conditions is the
 so-called problem of {\it universality}. Thus, universality features
 are linked to the degree of anisotropy and non-homogeneities of
 turbulent statistics \cite{bp.physrep}. Similar problems have already
 been explored measuring the velocity fluctuations in the laboratory
 frame (Eulerian statistics), where clear evidence of universality
 have been obtained \cite{ArETAL.96}.\\ To build a general theory of
 turbulent statistics, universality is the first requirement and, if
 proved, may open the possibility for effective stochastic
 modeling~\cite{Sawford91} in many applied situations.\\ This Letter
 aims at investigating intermittency and universality properties of
 velocity temporal fluctuations by quantitatively comparing data
 obtained from the most advanced laboratory~\cite{berg,MMMP.01,X06}
 and numerical~\cite{BBCLT.05,MLP04,mueller,pkyeung,chicago}
 experiments.  Main outcomes of our analysis are twofold. First, we
 show that data collapse on a common functional form, providing
 evidence for universality of velocity fluctuations --up to moments
 currently achievable with high statistical accuracy. At intermediate
 and inertial scales, data show an intermittent behaviour. Second, we
 propose a stochastic phenomenological modelisation in the entire
 range of scales, using a Multifractal description linking Eulerian
 and Lagrangian statistics.

We analyse the probability distribution of velocity fluctuations at
all scales, focusing on moments of these distributions, namely the
Lagrangian Velocity Structure Functions (LVSF) of positive integer
order $p$:
\begin{equation}
S_i^{(p)}(\tau) = \langle [v_i(t+\tau) -v_i(t)]^p \rangle=
\langle (\delta_\tau v_i)^p \rangle, 
\label{eq:LSF}
\end{equation} 
where $i=x,y,z$ are the velocity components along a single particle
path, and the average is defined over the ensemble of trajectories. As
stationarity and homogeneity is assumed, moments of velocity
increments only depend on the time lag $\tau$. In the inertial range,
for $\tau_\eta \ll \tau \ll T_L$, non-linear energy transfer governs
the dynamics.  Thus, from a dimensional viewpoint, only the scale
$\tau$ and the average energy dissipation rate for unit mass
$\epsilon$ should matter for the structure function behaviour.  The
only admissible choice is $S_i^{(p)}(\tau) \sim (\epsilon
\tau)^{p/2}$, but it does not take into account the fluctuating nature
of energy dissipation.  Empirical studies have indeed shown that the
tails of the probability density functions of $\delta_\tau v$ become
increasingly non-Gaussian at decreasing $\tau/T_L$. In terms of
  moments of the velocity fluctuations, intermittency reveals itself
  in the anomalous scaling exponents, i.e. a breakdown of the
  dimensional law for which we have that
\begin{equation}
S_i^{(p)}(\tau)  \sim \tau^{\xi(p)}\,, 
\label{eq:scaling}
\end{equation}
with $\xi(p)\neq p/2$.
Notice that when dissipative effects dominate, typically for scales
$\tau \sim \tau_{\eta}$ and smaller, the power-law behaviour
(\ref{eq:scaling}) is no longer valid, and refined arguments have to
be employed, as we will see in the following. 
\begin{table}[!t]
\begin{tabular}{|c|c|c|c|c|c|c|}
\hline
  EXP & $R_\lambda$ & $\tau_\eta$ ($s$)& meas. vol. ($\eta^3$) & $N_{tr}$   & Tech. & Ref.\\
\hline
  1   & 124          & $8.5\times 10^{-2}$&      $340^3$   & $1.6 \times 10^6$  & PTV  & [8] \\ 
   2   & 690          & $9 \times 10^{-4}$     &   $1700^3$ & $ 6.0 \times 10^6$  & PTV & [7] \\
 3   & 740          & $2\times 10^{-4}$      &  $6600^3$  & $9.5 \times 10^3$ & AD & [6] \\  
\hline
\end{tabular}
\caption{\label{table:exp} Experiments. By columns: 1- experiment
  label; 2- Taylor Reynolds number; 3- Kolmogorov time scale
  $\tau_\eta$; 4- measurement volume in unit of the Kolmogorov length
  scale $\eta$; 5- $N_{tr}$ total number of Lagrangian trajectories
  measured; 6- measurement technique: Particle Tracking Velocimetry
  (PTV) and Acoustic Doppler (AD); 7- Reference where information on
  the way the corresponding dataset was obtained can be found. }
\end{table}

\begin{table}[t!]
\begin{tabular}{|c|c|c|c|c|c|c|}
\hline
DNS & $R_\lambda$ & $N^3$ &  $N_{tr}$ & Diss. & Tech. & Ref.\\  \hline 
 1 & 140 & $256^3$ & $5\times 10^5$         & N  & T & [11]\\
 2 & 320 & $1024^3$ & $5\times 10^6$   & N  & T & [13]\\ 
 3 & 400 & $2048^3$ & $3\times 10^5$   & N  & L & [10]\\
 4 & 600 & $1856^3$ & $1.6\times 10^7$ & C  & L & [18]\\
 5 & 650 & $2048^3$ & $4\times 10^5$   & N  & CS & [12]\\
\hline
\end{tabular}
\caption{\label{tab:dns} Direct numerical simulations.  By columns: 1-
  numerical simulation label; 2- Taylor Reynolds number $R_\lambda$;
  3- number of collocation points $N^3$; 4- total number of Lagrangian
  tracers $N_p$; 5- characteristic of dissipation: normal viscous
  terms (N), weakly compressible code (C); 6- interpolation technique
  for Lagrangian integration: linear interpolation (L), tricubic
  interpolation (T); cubic-splines (CS); 7-  Reference where
  information on the way the corresponding dataset was obtained can be
  found.}
%\vspace{-0.8truecm}
\end{table}
The statistics of velocity fluctuations at varying time lag $\tau$ can
be quantitatively captured by the logarithmic derivatives of
$S_i^{(p)}(\tau)$ versus $S_i^{(2)}(\tau)$
\cite{ess,LM.04,PoF.07}. This defines the local scaling exponents
\begin{equation}
\zeta_i(p,\tau) = \frac{{\rm d} \log S_i^{(p)}(\tau)}{ {\rm d} \log S_i^{(2)}
  (\tau)}\,.
\label{eq:local}
\end{equation}
For statistically isotropic turbulence, all components are equivalent,
so that their spread quantifies the degree of anisotropy present in
teh flow. The $\tau$-dependence of $\zeta_i(p,\tau)$ allows for a
scale-by-scale characterisation of intermittency.
\begin{figure*}[t!]
\begin{center}
\includegraphics[width=0.7\textwidth]{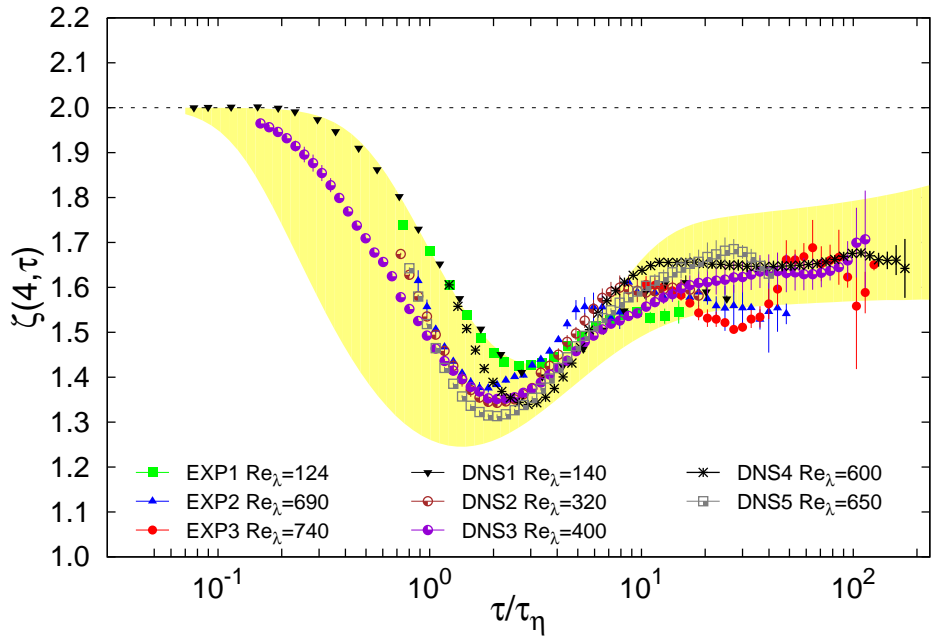}
\caption{(Color online) Log-Lin plot of the fourth order local
  exponent, $\zeta(4,\tau)$, averaged over velocity components, as a
  function of the normalised time lag $\tau/\tau_\eta$.  Data sets
  come from three experiments (EXP) (see Table 1) and five direct
  numerical simulations (DNS) (see Table 2).  Error bars are estimated
  from the spread between the three components, except in EXP3 where
  only two components were measured. Each data set is plotted only in
  the time range where recognised experimental/numerical limitations
  are not affecting the results. In particular, for each data set, the
  largest time lag always satisfies $\tau < T_L$. The minimal time lag
  is set by the highest fully resolved frequency.  The shaded area
  displays the prediction obtained by the MF model by using
  $D_{Lo}(h)$ or $D_{Tr}(h)$, with $\beta=4$, for a range of
  $R_\lambda \in [150:800]$, comparable with the range of $R_\lambda$
  in the data.  Notice that the MF predictions have been obtained by
  fixing equal to $7$, the multiplicative constant in the definition
  of $\tau_{\eta}$. The straight dashed line corresponds to the
  dimensional non-intermittent value $\zeta(4,\tau) =2$, achieved at
  small time lags where structure functions do become
  differentiable. Notice that two among the DNS are sufficiently
  resolved to get the mentioned dimensional scaling in the high
  frequency limit.\label{fig:1} }
\end{center}
\vspace{-0.3truecm}
\end{figure*}

Figure~1 shows the local exponents of order $p=4$ from a collection
of eight data sets, see Table I and II, for different Reynolds
numbers. Most of these data sets are new, as well as completely
new is the performed analysis, here presented. We focused on the
fourth order moment, since it is the highest order achievable with
statistical convergence for all data sets. Two observations can be
done. First, all data sets show a similar strong variation around the
dissipative time $\tau/\tau_{\eta} \sim O(1)$ that depends on the
Reynolds number, and then a clear tendency toward a plateau for larger
lags $\tau>10\tau_\eta$.  Second, all data sets, with comparable
Reynolds numbers, well agree in the whole range of time lags. The
relative scatter increases only for large $\tau$, due to the combined
effects of the lack of statistics, the anisotropy of the flows and the
different valus of $R_{\lambda}$. In particular, finite volume effects
in experimental particle tracking can produce a small -- but
systematic -- downward shift of the points at long-lag times
\cite{ott_mann,PoF.07}. It is worth noticing that error bars
estimated from anisotropic contributions decrease by going to small
$\tau$, indicating that isotropy tends to be recovered at sufficiently
small scales, i.e. large scale anisotropic contributions becomes less
and less important. In addition, the fact that, at comparable
Reynolds numbers, all data sets recover the same behaviour by going to
smaller and smaller time lags provides a clear indication of
Lagrangian {\it universality} of the energy cascade. Such an agreement
has not been observed before and is comparable with that found for the
corresponding Eulerian quantities \cite{ArETAL.96}.\\ The quality of
data shown in (Fig. 1) opens the possibility to quantitatively test
phenomenological models for LVSF, scale-by-scale.  Parisi-Frisch
Multifractal (MF) model of the inertial range~\cite{Fr.95}, and its
generalization to the dissipative range
\cite{PV.87,Ne.90,FV.91,meneveau}, has proved to give a satisfactory
description of Eulerian and Lagrangian fluctuations
\cite{Bo.93,bof02,CRLMPA.03,BBCDLT.04}. It is thus appealing to search
for a link between Eulerian and Lagrangian
statistics~\cite{Bo.93,bof02,CRLMPA.03,BBCDLT.04}, since this points
to a unique interpretation of turbulent fluctuations. Moreover, it
would reduce the number of free parameters. According to the MF model,
Eulerian velocity increments at inertial scales are characterised by a
local H\"older exponent $h$, i.e. $\delta_r u \sim r^h$, whose
probability ${\cal P}(h) \sim r^{3-D(h)}$ is weighted by the Eulerian
fractal dimension $D(h)$ of the set where $h$ is
observed~\cite{Fr.95}. The dimensional relation $\tau \sim r/\delta_r
u$ bridges Lagrangian fluctuations over a time lag $\tau$ to the
Eulerian ones at scale $r$. Following Refs.~\cite{Bo.93,meneveau}, it
is shown in Ref.~\cite{CRLMPA.03} how to extend the MF framework to
get a unified description at all time scales for Lagrangian
turbulence. Accordingly, Lagrangian increments display a continuous
and differentiable behaviour at the transition from the dissipative to
the inertial range,
\begin{equation}
\label{eq:fit}
\delta_{\tau} v(h) = V_0
\,\frac{\tau}{T_L}\left[\left(\frac{\tau}{T_L}\right)^{\beta} +
\left(\frac{\tau_{\eta}}{T_L}\right)^{\beta}\right]^{\frac{2h-1}{\beta(1-h)}}\,,
\end{equation}
$\beta$ being a free parameter controlling the crossover around $\tau\!
\sim\! \tau_{\eta}$, and $V_0$ the root mean square velocity. In order
to get a prediction for the behaviour of the LVSF, given by
\begin{equation}
\label{eq:lsfmulti}
\langle (\delta_{\tau} v)^p \rangle \sim  \int dh P_h(\tau,
\tau_{\eta}) [\delta_{\tau} v(h)]^p\,,
\end{equation}
we have to consider, in (\ref{eq:fit}), the intermittent fluctuations
of the dissipative scale~\cite{Bo.93,CRLMPA.03,BBCDLT.04},
$\tau_{\eta}(h)/T_L \!\sim\! R_{\lambda}^{2(h\!-\!1)/(1\!+\!h)}$. The
last necessary ingredient is to specify the probability of observing
fluctuations of $h$. This is done in analogy to Eq.~(\ref{eq:fit}):
\begin{equation}
P_h(\tau,\tau_{\eta}) = {\cal Z}^{-1}(\tau) \left[
  \left(\frac{\tau}{T_L}\right)^{\beta} +
  \left(\frac{\tau_{\eta}}{T_L}\right)^{\beta}\right]^{\frac{3-D(h)}{\beta(1-h)}},
\end{equation}
where ${\cal Z}$ is a normalizing function~\cite{CRLMPA.03} and $D(h)$
the fractal dimension of the support of the exponents $h$. Once
specified the Reynolds number, we are left with two parameters - the
expoent $\beta$ and a multiplicative constant in the definition of
$\tau_{\eta}$ -, while the function $D(h)$ comes from the knowledge of
the Eulerian statistics.\\
Eulerian Velocity Structure Functions (EVSF) have been measured in the
last two decades (see~Ref.\cite{ArETAL.96} for a data collection)
providing a way to estimate the function $D(h)$ based on empirical
data. Many functional forms have been proposed in the
literature~\cite{Fr.95} that are consistent with data, up to
statistical uncertainties. Eulerian velocity statistics can be
measured in terms of longitudinal or transverse fluctuations. Fluid
velocity along particle paths is naturally sensitive to both kinds of
fluctuations. We thus evaluated the LVSF in (\ref{eq:lsfmulti}) using
the fractal dimensions $D_{Lo}(h)$ and $D_{Tr}(h)$ obtained by
longitudinal~\cite{ArETAL.96} and transverse~\cite{Go.02} moments of
Eulerian luctuations, respectively.

The shaded area in (Fig. 1) represents the range of variation of the
MF prediction computed from $D_{Lo}(h)$ or $D_{Tr}(h)$, measured in
the Eulerian statistics (see below), and at changing Reynolds numbers.
This must be interpreted as our uncertainty. The prediction works very
well: all data fall within the shaded area. The role of the parameters
is clear. Changing $\beta$ modifies the sharpness and shape of the dip
region at $\tau_{\eta}$ -- the larger $\beta$ the more pronounced the
dip; while, changing the multiplicative constant in the definition of
$\tau_{\eta}$ has no effect on the curve shape, but it rigidly shifts
the whole curve along the time axis. \\Increasing the Reynolds number
$R_\lambda$, the flat region at large lags develops a longer
plateau. In the limit $R_\lambda\to \infty$ the MF model predicts
$\zeta(4) \simeq 1.71 $ from $D_{Lo}(h)$ and $\zeta(4) \simeq 1.59$
from $D_{Tr}(h)$ statistics.

For the Eulerian $D(h)$, we used the following log-Poisson~\cite{Fr.95}
functional form,
\begin{equation} 
D(h) = \frac{3 (h-h^*)}{ \log(\gamma)} \left[\log \left( \frac{3
    (h^*-h)}{d^* \log(\gamma)}\right) -1\right] +3 -d^*\,.
\end{equation}
Different couples of parameters, $(h^*,\gamma)$, have been chosen to
fit longitudinal and transverse Eulerian fluctuations. The parameter
$d^* = (1-3h^*)/(1-\gamma)$ is fixed by imposing the exact relation
for third order EVSF. For the longitudinal exponents \cite{ArETAL.96},
we used $(h^*_{Lo}=1/9,\gamma_{Lo}=2/3)$ \cite{Fr.95}. For the
transverse exponents, we used $(h^*_{Tr}=1/9,\gamma_{Tr}=1/2)$ which
fits the data in Ref.~\cite{Go.02} (see \cite{nota} for details.)

This comprehensive comparison of the best available experiments and
direct numerical simulations provides strong evidence of the
universality of Lagrangian statistics.  One important open question is
the effect of a mean flow, as in turbulent jets~\cite{Baudet} and wall
bounded turbulence, where strong persistence of anisotropy may break
the recovery of small-scale universality. We showed that a
Multifractal description is in good agreement with data, even in the
dissipative range where intermittency is significantly increased. The
Multifractal description captures the intermittency at all scales with
only a few parameters, independent of the Reynolds number. This is the
universal feature of Lagrangian turbulence revealed by this study.
There exists a long debate on the statistical importance of vortex
filaments around dissipative time and length scales
\cite{DCB.91,Fr.95}. Simulations~\cite{BBCLT.06,BBCLT.05,LM.04} show
that the dip region for $\tau \sim \tau_{\eta}$ can be
depleted/enhanced by decreasing/increasing the probability of
particles being trapped in vortex filaments. The Multifractal model is
able to capture the intermittency around $\tau_{\eta}$ with the help
of the free parameter $\beta$.  Different values of $\beta$ should
then correspond to different statistical weights of vortex filaments
along particle trajectories.\\ Only further advances in both
experimental techniques and numerical power will allow us to test the
same questions here addressed also for the higher order statistics.\\

Helpful discussions with U. Frisch and L.~P. Kadanoff are gratefully
acknowledged. F.T. thanks the DEISA Consortium (co-funded by the EU),
for support within the DEISA Extreme Computing Initiative; L.B., M.C.,
A.S.L. and F.T. thank CINECA (Bologna, Italy) for technical
support.\\$^* ictr.rome@roma2.infn.it$


\begin{thebibliography}{99}
\bibitem{nota} The functions $D_{Lo}(h)$ and $D_{Tr}(h)$ are
  convex; the strongest fluctuations, corresponding to the smallest
  H\"older exponent $h$, are realised for $h=h^*_{Lo}$ and $h^*_{Tr}$,
  respectively. The maxima of fractal dimensions,
  $D_{Lo}(h^{max}_{Lo})=D_{Tr}(h^{max}_{Tr})=3$, are attained for
  $h^{max}_{Lo}=0.382$, and $h^{max}_{Tr}=0.57$.  Since only the range
  $h^{*}\leq h \leq h^{max}$ is relevant to the fit of EVSF of
  positive order~\cite{Fr.95}, the extremes of integration in
  Eq.~(\ref{eq:lsfmulti}) have been set in the interval,
  $[h^*_{Lo},h^{max}_{Lo}]$ and $[h^*_{Tr},h^{max}_{Tr}]$,
  respectively. Modifying $h^{max}$ alters the MF prediction for $\tau
  \sim T_L$, but does not change the behaviour of the curves in the
  inertial range.
\end{thebibliography}
\end{document}